%% file: paper.tex
\PassOptionsToPackage{capitalize}{cleveref}
\documentclass[twocolumn]{fairmeta}

\input{others/packages}
\input{others/commands}

\title{\thiswork: Joint Draft and Verification Budget Scheduling for Parallel Speculative Decoding}

\author[1,2,*]{Jaeyong Song}
\author[1]{Daniel Wang}
\author[1]{Pratyush Patel}
\author[1]{Jonny Kong}
\author[1]{Preyas Shah}
\author[1]{Shobhit Kanaujia}

\affiliation[1]{Meta}
\affiliation[2]{Seoul National University}

\contribution[*]{Work done during an internship at Meta}

\abstract{\input{sections/0_abstract}}

\date{\today}
\correspondence{Jaeyong Song at \email{jaeyong@meta.com} or \email{jaeyong.song@snu.ac.kr}}

\begin{document}

\maketitle

\input{sections/1_introduction}
\input{sections/2_background}
\input{sections/3_motivation}
\input{sections/4_design}
\input{sections/5_implementation}
\input{sections/6_evaluation}
\input{sections/7_discussion}

\input{sections/8_related_work}
\input{sections/9_conclusion}

\bibliographystyle{assets/plainnat}
\bibliography{references}

\input{sections/etc_ai_usage_disclosure}

\clearpage
\beginappendix
\crefalias{section}{appendix}

\input{sections/etc_appendix}

\end{document}

%% file: others/packages.tex
\usepackage{amsmath}
\usepackage{amssymb}

\usepackage{xurl}

\usepackage{xspace}

\usepackage{enumitem}

\let\cite\citep

\DeclareFontShape{T1}{optimistic}{m}{sc}{<-> s * [0.88] assets/optimistic}{}
\DeclareFontShape{T1}{optimistic}{b}{sc}{<-> s * [0.88] assets/optimistic}{}

\titleformat*{\section}{\Large\sffamily\bfseries\filright}
\titleformat*{\subsection}{\large\sffamily\bfseries\filright}

%% file: others/commands.tex
\newcommand{\thiswork}{\textsc{ForeSpark}\xspace}
\newcommand{\JY}[1]{}
\newcommand{\DW}[1]{}
\newcommand{\PP}[1]{}
\newcommand{\JK}[1]{}

\newcommand{\adaptveri}{adaptive verification\xspace}

\newcommand{\ourconcept}{work-conserving speculation\xspace}

\newcommand{\predraft}{pre-draft estimate\xspace}
\newcommand{\Predraft}{Pre-draft estimate\xspace}

\newcommand{\jointschedule}{joint schedule\xspace}
\newcommand{\Jointschedule}{Joint schedule\xspace}

\newcommand{\postrefine}{distribute\xspace}
\newcommand{\Postrefine}{Distribute\xspace}

\newcommand{\dspark}{\textsc{DSpark}\xspace}
\newcommand{\dflashp}{\textsc{DFlash2}\xspace}
\newcommand{\dflashpp}{\textsc{BW}\xspace}
\newcommand{\oursnojs}{\textsc{ForeSpark} w/o JS\xspace}

\newcommand{\mtp}{\textsc{MTP}\xspace}
\newcommand{\ar}{\textsc{AR}\xspace}
\newcommand{\eagle}{\textsc{EAGLE-3}\xspace}

\makeatletter
\DeclareRobustCommand{\@modelsep}{\@ifnextchar/{\@modelslash}{\xspace}}
\def\@modelslash/{\slash}
\newcommand{\qwen}{\texttt{Qwen3-14B}\@modelsep}
\newcommand{\glmflash}{\texttt{GLM5.3-Flash}\@modelsep}
\newcommand{\glm}{\texttt{GLM5.3}\@modelsep}
\makeatother

\usepackage{tikz}

\newcommand*\circled[1]{\tikz[baseline=(char.base)]{
            \node[shape=circle,draw,inner sep=0.4pt, fill=white, text=black] (char) {#1};}}

\newcommand*{\doi}[1]{doi:~\href{https://doi.org/#1}{\nolinkurl{#1}}}

%% file: sections/0_abstract.tex
Parallel drafters accelerate LLM decoding by proposing a whole block of draft tokens in a single forward pass, making them attractive for LLM serving.
Recent systems optimize goodput (i.e., accepted tokens per second) because rejected drafts steal compute from other requests.
However, these systems control only how many tokens are verified, so every request still drafts a full block.
At high serving load, we find that drafting consumes up to 40\% of step time, of which a majority is discarded because it is never verified.
Yet, draft-length control from autoregressive drafters does not carry over, because a parallel drafter commits to its length before its single pass.
Moreover, draft and verification lengths are coupled across the batch, so they must be selected jointly for every request while keeping the decision computationally cheap at every step.
In this paper, we propose \textit{\ourconcept (WCS)}, a design that jointly allocates draft and verification budgets per request in LLM serving, and present \thiswork as its prototype.
\thiswork realizes WCS with (i) a context-aware pre-draft estimator that predicts acceptance for every candidate length from the target model's context features, (ii) a lightweight joint draft-verify scheduler that selects both lengths through a global sort, and (iii) a post-draft budget distribution that spends verification on the most promising tokens.
Implemented on vLLM, \thiswork improves throughput by up to 1.77$\times$ over the state-of-the-art \dspark and by 2.49$\times$ over autoregressive decoding at a matched per-request generation rate, without changing the output distribution.
It also tolerates higher load, sustaining up to 1.51$\times$ the request rate of \dspark under latency service level objectives (SLOs).

%% file: sections/1_introduction.tex
\section{Introduction}
\label{sec:intro}

Speculative decoding leverages the idle compute during LLM decoding to accelerate inference~\cite{leviathan23,magicdec}.
A drafter proposes several tokens that the target model verifies in one forward pass without changing its output distribution~\cite{leviathan23,eagle,dflash}.

To avoid the per-token forward passes of autoregressive drafters (e.g., MTP~\cite{deepseekv3} and EAGLE~\cite{eagle,eagle2,eagle3}), \textit{parallel drafters}~\cite{dflash,dflash2,dspark} propose a block of $K$ tokens in one forward pass.
Within the block, the drafter typically attends bidirectionally, so the logits for all $K$ positions are produced in one backbone pass.

\begin{figure}[t]
    \centering
    \includegraphics[width=\columnwidth]{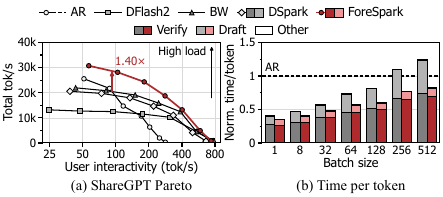}
    \caption{\qwen on ShareGPT-E2E.
    (a) Total throughput versus user interactivity (per-user throughput), where \dspark~\cite{dspark} outperforms \dflashp with adaptive verification and \thiswork adds joint scheduling.
    (b) Time per output token of \dspark and \thiswork, normalized to \ar (i.e., no speculation). See \cref{sec:eval:setup} for the detailed setup. }
    \label{fig:intro}
\end{figure}

Recent parallel drafters~\cite{dspark,adaflash} optimize goodput, the number of accepted tokens per second, because under high load, verification becomes compute-bound and rejected draft tokens take compute from other requests~\cite{magicdec,turbospec}.
They address this through \textit{\adaptveri}, which uses predicted acceptance to truncate each request's verification to its most promising drafts.
As \cref{fig:intro}(a) shows, adaptive verification (\dspark) substantially lifts throughput over full verification (\dflashp) under high load.

Adaptive verification, however, still \textit{drafts a full block of $K$ tokens for every request}, implicitly assuming the parallel draft is computationally cheap.
We find that this assumption is costly under high load.
As \cref{fig:intro}(b) shows, drafting takes up to 40\% of \dspark's time per output token (gray bars).
The reason is a mismatch between drafting and verification: at a load of 512 requests, adaptive verification retains only 3.77 of the 15 drafted tokens per request, while the drafter still generates all 15.
Thus, 75\% of the drafter's computation is discarded, leaving LLM serving suboptimal.

A simple remedy is vLLM's load-based batch-wise length cap (\dflashpp in \cref{fig:intro}(a)), which shortens every speculation under load and already edges out \dspark at several loads.
Yet it applies one length to every request, so a better remedy is to choose the length per request, adapting approaches for autoregressive drafters~\cite{specdecpp,adaeagle}.
However, parallel drafters differ structurally from autoregressive ones, which poses two challenges.

\begin{enumerate}[leftmargin=*]

\item \textit{Draft length cannot be adjusted during drafting.}
Autoregressive drafters evaluate draft quality token by token and can stop as soon as their confidence drops~\cite{specdecpp}.
Parallel drafters instead commit to a block size before their single pass, so they cannot prune candidates during drafting, and adaptive verification estimates acceptance only after the drafting compute is spent~\cite{dspark,adaflash}.
Past acceptance~\cite{diffuspec} is a poor proxy~\cite{failfast}, changing across consecutive steps in over 80\% of cases (\cref{sec:motivation}).
Adaptive verification also makes this proxy unexploitable, as truncation hides the actual acceptance length.

\item \textit{Joint length allocation is combinatorial yet must be computationally cheap.}
The scheduler must jointly choose, per request, a draft length and a verification budget within one step, while keeping the decision computationally cheap.
The two lengths are coupled, making the decision combinatorial, since a token never drafted cannot be verified while a surplus draft token can be dropped during verification, so the scheduler must sometimes over-draft and set both lengths together to maximize goodput.

\end{enumerate}

To address these challenges, we propose \emph{\ourconcept (WCS)}, a design for jointly allocating draft and verification budgets per request in LLM serving, and build \thiswork as its prototype.
WCS organizes each draft-and-verify step into four coordinated stages, \textit{\predraft}, \textit{\jointschedule}, \textit{draft}, and \textit{\postrefine}, so that drafting and verification compute is spent only where it pays off, the sense in which we call it work-conserving.
\thiswork realizes WCS through the following methods.

\begin{itemize}[leftmargin=*]
\item \textit{Context-aware pre-draft estimation.}
We observe that the target's context features at the last accepted token signal how predictable upcoming tokens are.
Before drafting, a lightweight estimator reads them to predict acceptance for every candidate length, helping address both challenges.

\item \textit{Lightweight joint scheduling.}
To address the combinatorial search overhead, we replace each request's expected acceptance across candidate lengths with its concave envelope.
This reduces multi-request length allocation to one global sort, so the scheduler assigns per-request draft lengths and verification budgets.
When drafting is computationally cheap, the scheduler intentionally over-drafts and defers the verification decision to the post-draft stage.

\item \textit{Post-draft budget distribution.}
A variable-length drafter executes the schedule as one packed batch, so shorter drafts save compute directly.
A post-draft estimator re-scores the drafts, and the system distributes the verification budget across the batch to the most promising ones.
\end{itemize}

We implement \thiswork on vLLM~0.30.0~\cite{vllm} and evaluate it on \qwen, \glmflash, and \glm, which span a range of model types and sizes.
\thiswork improves throughput by up to 1.77$\times$ over the state-of-the-art \dspark and by 2.49$\times$ over autoregressive decoding at a matched per-request generation rate, without changing the output distribution.
As \cref{fig:intro} previews, \thiswork attains the highest throughput by jointly scheduling drafting and verification.
It also tolerates higher load, sustaining up to 1.51$\times$ the request rate of \dspark under latency SLOs.
Because full-block drafting with adaptive verification, as in \dspark, is a special case within \thiswork's design space, \thiswork performs comparably to \dspark under low load.
Since WCS is orthogonal to drafter design, it can also lift the throughput of existing parallel and autoregressive drafters (\cref{sec:eval:other_drafters}).
We believe WCS provides a foundation for future parallel drafters in LLM serving.

%% file: sections/2_background.tex
\section{Background}
\label{sec:background}

\begin{figure}[t]
    \centering
    \includegraphics[width=\columnwidth]{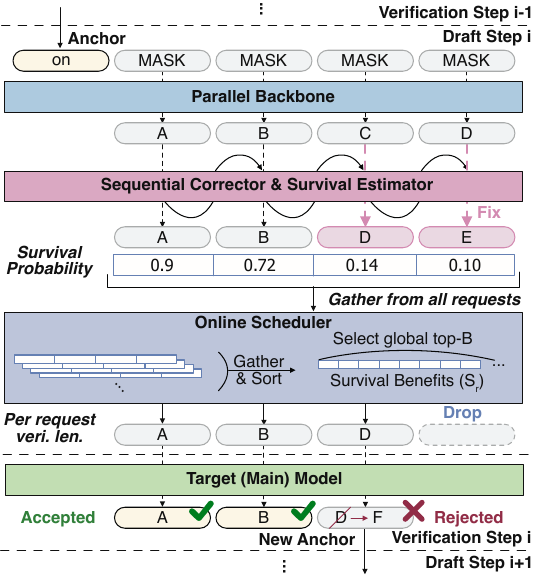}
    \caption{Parallel speculative decoding example with \dspark.}
    \label{fig:background}
\end{figure}

\textbf{Speculative decoding.}
Speculative decoding generates multiple tokens per target-model pass by pairing the target with a lightweight drafter~\cite{leviathan23}.
As depicted in \cref{fig:background}, in each step the drafter extends the \textit{anchor}, the last accepted or generated token, by up to \(K\) proposed tokens (\(K\)=4 in the example).
The target model verifies all proposals in one forward pass, accepts the longest matching prefix (e.g., \texttt{A,B}), and emits one bonus token (e.g., \texttt{F}) that becomes the next anchor.
With \(\tau\) tokens emitted per step, throughput is \(\tau/(T_{\mathrm{draft}}+T_{\mathrm{verify}})\), so speculation pays off only when \(\tau\) is high and both times stay small.

\textbf{Parallel speculative decoding.}
Autoregressive drafters~\cite{eagle,eagle2,eagle3,deepseekv3} run one pass per draft position, so \(T_{\mathrm{draft}}\) grows with \(K\).
As shown in \cref{fig:background}, parallel drafters instead fill a block of \(K\) mask tokens and generate all \(K\) tokens in a single pass of a \textit{parallel backbone} that attends to target-model features~\cite{dflash}.
Within a block, bidirectional attention modestly raises acceptance over causal attention~\cite{adaflash}, but as no position sees the tokens chosen before it, later proposals can contradict earlier ones.
\dspark repairs this with a sequential corrector that conditions each position on its predecessor (\texttt{C}$\xrightarrow{}$\texttt{D}, \texttt{D}$\xrightarrow{}$\texttt{E})~\cite{dspark}, and similarly, \dflashp addresses this with a lightweight path selector over the backbone's top candidates~\cite{dflash2}.

\textbf{Goodput-aware speculation.}
A serving engine batches \(n\) requests, and each verification step processes \(N\) non-speculative tokens (one anchor per decoding request plus prefill tokens) and \(B\) draft tokens.
Beyond some load, verification becomes compute-bound and each extra verified token lengthens the whole batch's step, so serving systems consider \textit{goodput}, the rate of accepted tokens, alongside acceptance~\cite{magicdec,turbospec}.

\dspark pursues goodput with \textit{adaptive verification}~\cite{dspark}.
For request \(i\), its corrector estimates the probability \(\omega_{i,j}\) that the token at draft position \(j\) (the \(j\)-th token after the anchor) is accepted given its predecessors, and the survival probability \(S_{i,j}=\prod_{t\le j}\omega_{i,t}\) gives the expected gain of verifying position \(j\) (0.9, 0.72, 0.14, and 0.10 in \cref{fig:background}).
The scheduler pools \(S_{i,j}\) from all requests, sorts them so that \(S_{(r)}\) is the \(r\)-th largest, and picks the budget
\begin{equation}
\operatorname*{arg\,max}_{B}\frac{n+\sum_{r=1}^{B}S_{(r)}}{T_{\mathrm{draft}}+T_{\mathrm{verify}}(N+B)},
\label{eq:dspark_budget}
\end{equation}
where the numerator counts one guaranteed token per request plus the expected accepted drafts, and the denominator is the profiled step latency.
Since survival only shrinks along a draft, the selection is always a prefix of each draft (\texttt{A,B,D}, dropping \texttt{E}), which verification requires.
AdaFlash similarly truncates verification based on a predicted acceptance ratio~\cite{adaflash}.
Adaptive verification, however, still drafts all \(K\) tokens before deciding how many to verify.

%% file: sections/3_motivation.tex
\section{Motivational Study}
\label{sec:motivation}

\textbf{Drafter overhead under adaptive verification.}
Adaptive verification shrinks the target's verification work under load, but the drafter still produces all \(K\) tokens for every request, so drafting grows to up to 40\% of step time (\cref{fig:intro}(b)).
This also limits the adoption of a larger \(K\), which typically yields longer acceptance~\cite{dflash} but further raises drafting cost under load (\cref{app:k_7}).
Borrowing draft-length control from prior work appears to be a natural fix, but two challenges stand in the way.

\begin{figure}[t]
    \centering
    \includegraphics[width=\columnwidth]{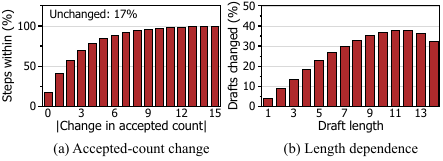}
    \caption{\dflashp~\cite{dflash2} on \qwen ($K$=15) on ShareGPT.
    (a) Step-to-step accepted-count change.
    (b) Share of $d$-token drafts differing from the full draft's first $d$ tokens.}
    \label{fig:moti}
\end{figure}

\textbf{Draft length cannot be adjusted during drafting.}
Unlike an autoregressive drafter, which can stop once its confidence drops~\cite{specdecpp}, a parallel drafter produces all \(K\) tokens in one pass, so it must fix its length before any draft token or confidence exists.
The length must therefore be estimated beforehand, and prior work uses past acceptance for this~\cite{diffuspec}.
Past acceptance is a poor proxy~\cite{failfast}, as the accepted count changes between consecutive steps in over 80\% of cases and by three or more tokens in about 40\% (\cref{fig:moti}(a)).
Adaptive verification also hides it, since tokens beyond the verified prefix are never scored, so a history-based controller that shortens a draft observes only shorter acceptance and may keep shrinking it.
A load-aware, batch-wise \(K\) (\cref{fig:intro}(a), \dflashpp) avoids estimation, but acceptance varies across requests as well as across steps, so one length for the whole batch cannot follow it.

\textbf{Joint length allocation is combinatorial yet must be computationally cheap.}
\dspark's objective treats drafting latency as a constant (\cref{eq:dspark_budget}), so it never trades drafted tokens against drafting time.
Making each draft length a decision variable couples it with the verification budget, because a token never drafted cannot be verified, and the joint space grows as $K^n$ for $n$ requests while the choice must be made at every step.
Bidirectional attention adds another complication.
For each length \(d\), \cref{fig:moti}(b) shows that 4--38\% of \(d\)-token drafts differ in at least one token from the first \(d\) tokens of the 15-token draft, so a shorter draft is a different proposal rather than a truncation.
Thus, a scheduler should consider every draft length ($d \le K$) separately.\footnote{Although these differences are large, on the \dflashp backbone a single length-$K$ curve predicts acceptance as accurately as per-length estimates (\cref{sec:eval:ablation}). \thiswork still handles the general case at a similar cost for drafters where this may not hold.}

These observations motivate \ourconcept, which estimates per-length acceptance before drafting and allocates draft and verification lengths jointly per request.

%% file: sections/4_design.tex
\section{\thiswork: Realizing Work-Conserving Speculation}
\label{sec:design}

\begin{figure}[t]
    \centering
    \includegraphics[width=\columnwidth]{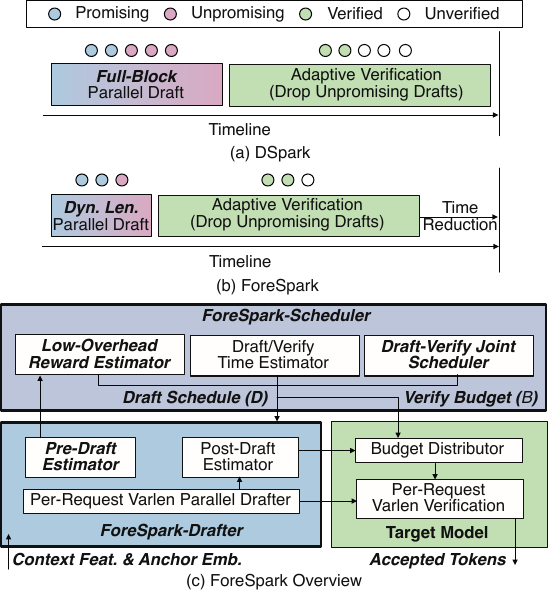}
    \caption{(a) \dspark drafts a full block and drops drafts only at verification.
    (b) \thiswork sizes each draft before drafting.
    (c) Overview of \thiswork.}
    \label{fig:overview}
\end{figure}

\thiswork realizes \ourconcept by deciding how much to draft before any drafting cost is paid.
As \cref{fig:overview}(a,b) contrasts, \dspark drafts a full block and drops unpromising drafts only at verification, whereas \thiswork sizes each draft beforehand, so far fewer drafted tokens go unverified.
\cref{fig:overview}(c) maps the four stages of WCS onto \thiswork system components.
\circled{1} \textit{\Predraft}. The pre-draft estimator predicts each request's acceptance at every candidate length (\cref{sec:predraft}).
\circled{2} \textit{\Jointschedule}. The \thiswork-Scheduler turns these predictions into a per-request draft schedule \(\mathbf{D}=(d_1,\ldots,d_n)\) and a verification budget \(B\) (\cref{sec:scheduler}).
\circled{3} \textit{Draft}. The variable-length drafter executes \(\mathbf{D}\).
\circled{4} \textit{\Postrefine}. The post-draft estimator re-scores the drafted tokens, and the budget distributor allocates \(B\) to the most promising ones (\cref{sec:postdraft}).
Because only the total budget \(B\) is fixed before drafting, \thiswork distributes it per request using the actually realized drafts.
Leaving this distribution to the GPU also reduces host-GPU synchronization (\cref{sec:implementation}).

\subsection{Preliminaries: Parallel Drafting Backbone}
\label{sec:prelim}

We build on \dflashp~\cite{dflash2}, which forms the drafter in \cref{fig:overview}(c), because it attains acceptance lengths similar to \dspark (\cref{sec:eval:qualitative}) while replacing the sequential corrector with a lightweight path selector over the backbone's top candidates.
Following EAGLE-3~\cite{eagle3}, the drafter conditions on the target model's hidden states from several intermediate layers.
A learned projection fuses those hidden states into context features $H_{\mathrm{feat}}\in\mathbb{R}^{T\times d_h}$, one vector for each of the $T$ tokens in the target's context, where $d_h$ is the drafter's hidden dimension.
For request $i$, a length-$d$ draft takes as input the embedding of the anchor token followed by $d$ learned mask embeddings, one for each draft position $j=1,\ldots,d$, similar to \cref{fig:background}.
Each draft layer attends bidirectionally to all draft positions and to $H_{\mathrm{feat}}$, which enters only as additional keys and values.
The last layer yields a hidden state $h_{i,j}$ for each draft position $j$, and the LM head turns it into the proposed token $x_{i,j}$.

\subsection{Context-Aware Pre-Draft Estimation}
\label{sec:predraft}

\begin{figure}[t]
    \centering
    \includegraphics[width=\columnwidth]{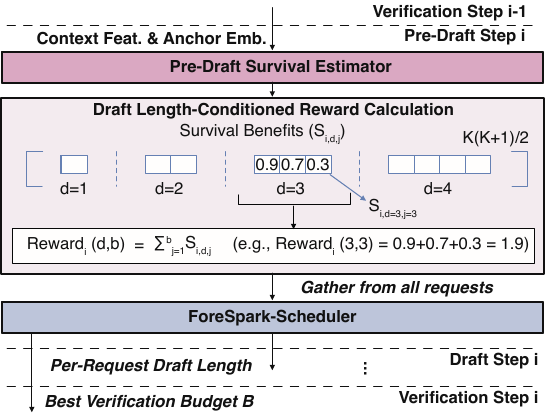}
    \caption{Pre-draft estimation.
    The estimator predicts survival benefits for every candidate length, a triangle of $K(K+1)/2$ values, from which the scheduler gathers per-length rewards.}
    \label{fig:predraft}
\end{figure}

As \cref{fig:predraft} shows, the \textit{pre-draft survival estimator} predicts, before drafting, how every candidate draft length would perform for each request.
Instead of past acceptance (\cref{sec:motivation}), it reads two inputs from the verification pass that just finished, both describing what comes next: the context feature $c_i=H_{\mathrm{feat}}[a_i]$, the row of $H_{\mathrm{feat}}$ at the anchor's position $a_i$ in the context, and the anchor token's embedding $e_i$.
Since a length-$d$ draft need not be the prefix of a longer one, the estimator predicts a separate survival curve for every candidate length.
To stay computationally cheap on the critical path between verification and drafting, it fuses the inputs in a low-rank space of dimension $r\ll d_h$ ($r$=256 by default) with one hidden layer,
\begin{equation}
\begin{aligned}
u_i&=\operatorname{SiLU}\left(W_c c_i+W_e e_i\right)\in\mathbb{R}^{r},\\
z_i&=W_o u_i\in\mathbb{R}^{K(K+1)/2},
\end{aligned}
\end{equation}
where $W_c$ and $W_e$ project $c_i$ and $e_i$ into a shared $r$-dimensional hidden representation $u_i$, and $W_o$ maps $u_i$ to the $K(K+1)/2$ conditional-acceptance logits, with biases omitted for clarity.
Applying SiLU to $W_c c_i+W_e e_i$ fuses the two inputs nonlinearly, and $W_o$ reads every candidate length from $u_i$.
As $c_i$ and $e_i$ are already rich target representations, a single hidden layer suffices, and the estimator costs only a few matrix-vector products per request.

As depicted in \cref{fig:predraft}, the output forms a triangle whose row $z_i^{(d)}\in\mathbb{R}^{d}$ describes a length-$d$ draft, with entry $z_{i,t}^{(d)}$ being the conditional-acceptance logit of position $t$, and the running product of their sigmoids gives the survival benefit
\begin{equation}
S_{i,d,j}=\prod_{t=1}^{j}\sigma\big(z_{i,t}^{(d)}\big)\approx\Pr\!\left(A_{i,d}\geq j\right),
\end{equation}
where $\sigma$ is the sigmoid function and $A_{i,d}$ is the number of accepted tokens when request $i$ drafts $d$ tokens.
Verifying the first $b\leq d$ tokens of a length-$d$ draft is then expected to yield $\mathrm{Reward}_i(d,b)=\sum_{j=1}^{b}S_{i,d,j}$ accepted tokens, so the length-3 draft in \cref{fig:predraft} with benefits $0.9,0.7,0.3$ gives $\mathrm{Reward}_i(3,3)=1.9$.
The estimator predicts per-position acceptance rather than the scheduler's survival scores (\cref{sec:scheduler}), since verification already labels acceptance for training (\cref{sec:training}).
For a \textit{directional} drafter, i.e., one with causal attention within the block, shorter drafts are exact prefixes of longer ones, so every row equals a prefix of row $K$ and $W_o$ needs only $K$ logits instead of $K(K+1)/2$.
\thiswork also works well for such a drafter (\cref{sec:eval:sensitivity}).

\subsection{Lightweight Draft-Verify Joint Scheduling}
\label{sec:scheduler}

\begin{figure}[t]
    \centering
    \includegraphics[width=\columnwidth]{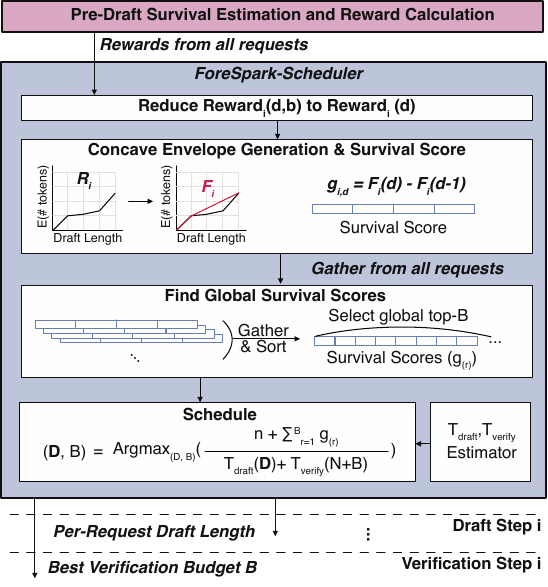}
    \caption{\thiswork-Scheduler.
    Survival scores, differences of concave envelopes $F_i$ over per-length rewards $R_i$, are pooled, sorted, and priced against latency profiles to choose $\mathbf{D}$ and $B$.}
    \label{fig:scheduler}
\end{figure}

As \cref{fig:scheduler} shows, the \thiswork-Scheduler turns per-length rewards into a draft schedule $\mathbf{D}$ and a verification budget $B$ in one pass.
Unlike \cref{eq:dspark_budget}, it prices drafting latency, so it drafts tightly at high load and generously at low load, where extra drafts add little latency.
Its search jointly covers every draft length of every request and the verification budget, yet stays computationally cheap as follows.

\textbf{Per-request reward curve generation.}
To choose draft lengths, the scheduler values each length-$d$ draft as if all its tokens were verified, $R_i(d)=\mathrm{Reward}_i(d,d)$ with $R_i(0)=0$, and leaves how many tokens to verify to the joint schedule later.
A longer draft may lift the yield of its earlier positions (\cref{fig:scheduler}, $R_i$ plot), so $R_i$ need not be concave, and a greedy rule that extends a draft only while the next token pays off could stop early and lose the larger gain of a longer draft.

\textbf{Concave envelope and survival score generation.}
Therefore, as shown in \cref{fig:scheduler}, the scheduler replaces $R_i$ with its upper concave envelope $F_i$, the smallest concave function above $R_i$ on $\{0,\ldots,K\}$, and defines the survival score $g_{i,d}=F_i(d)-F_i(d-1)$, the marginal gain of drafting the $d$-th token.
By concavity, the scores $g_{i,d}$ are non-increasing in $d$, which the next stage relies on.
Where an extra token alone pays off little but enables a larger gain later, $F_i$ spreads that gain evenly over the tokens in between, so the selection does not stop early.
Building the envelope takes $O(K)$ time and leaves a concave $R_i$ unchanged.

\textbf{Global survival score gather.}
As \dspark~\cite{dspark} does in \adaptveri, the scheduler pools all requests' scores and sorts them, $g_{(1)}\geq g_{(2)}\geq\cdots$, where $g_{(r)}$ is the $r$-th largest.
Because each request's scores are non-increasing, the top entries form a prefix of every request and directly yield a valid schedule $\mathbf{D}$ in the next step (i.e., the joint schedule).
This selection maximizes $\sum_i F_i(d_i)$ without searching $K^n$ schedules, averaging within 0.01\% of the optimum on real batches (\cref{sec:eval:ablation}).

\textbf{Joint schedule.}
Finally, the scheduler chooses $\mathbf{D}$ and $B$, verifying no more than it drafts,
\begin{equation}
(\mathbf{D},B)=\operatorname*{arg\,max}_{\mathbf{D},\,B\leq\sum_i d_i}\frac{n+\sum_{r=1}^{B}g_{(r)}}{T_{\mathrm{draft}}(\mathbf{D})+T_{\mathrm{verify}}(N+B)},
\label{eq:joint}
\end{equation}
where the numerator counts one token per request plus the expected accepted drafts among the top-$B$ positions, and the denominator is the offline-profiled step latency.
Since the scheduler fixes only the total $B$, which may stay below $\sum_i d_i$, it can over-draft and leave the per-request verification split to the post-draft stage (\cref{sec:postdraft}).
The search is computationally cheap as the drafter and target model replay CUDA graphs for a few token-count buckets, so each latency is constant within a bucket.
Within each pair of buckets, the largest feasible schedule is best, so the scheduler checks one candidate per pair, reducing \cref{eq:joint} to a small grid.

\subsection{Draft and Distribute: Variable-Length Drafting and Post-Draft Budget Distribution}
\label{sec:postdraft}

\begin{figure}[t]
    \centering
    \includegraphics[width=\columnwidth]{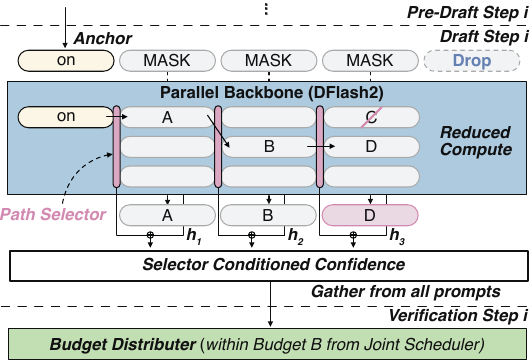}
    \caption{Variable-length drafting and budget distribution.
    Masks beyond the scheduled length are skipped, and post-draft survival estimates allocate the budget $B$.}
    \label{fig:postdraft}
\end{figure}

Once the drafts exist, the post-draft estimator can use information the pre-draft estimator never had (\cref{fig:postdraft}).
The pre-draft estimator decides how much drafting compute to commit, while the post-draft estimator sees the realized tokens and decides where the verification budget is best spent, which is what makes over-drafting worthwhile.

\textbf{Per-request variable-length drafting.}
\thiswork packs the blocks of all requests, each with its anchor and $d_i$ masks, into one flat batch with block-diagonal attention, so drafter compute scales with $\sum_i(d_i+1)$ rather than $n(K+1)$, and shorter drafts actually reduce drafting cost.
In \cref{fig:postdraft}, the fourth mask is dropped before the backbone runs, so it executes exactly the length the scheduler priced.

\textbf{Post-draft survival estimation.}
\dflashp's path selector keeps a few candidate tokens per position from the backbone and picks one by conditioning on the token chosen at the previous position~\cite{dflash2}, replacing \texttt{C} with \texttt{D} in the example.
The hidden state $h_{i,j}$ therefore reflects the backbone's belief before this choice, not the token $x_{i,j}$ actually proposed.
Similar to the confidence head of \dspark~\cite{dspark}, which pairs the backbone hidden state with an embedding of the previous token, the post-draft estimator pairs $h_{i,j}$ with a selector feature $s_{i,j}=(W_s h_{i,j})\odot E_{\mathrm{pred}}(x_{i,j-1})$, where $W_s$ is a learned projection, $E_{\mathrm{pred}}$ the selector's embedding of the previous token $x_{i,j-1}$, and $\odot$ the elementwise product,
\begin{equation}
p_{i,j}=\sigma\left(w^{\top}[h_{i,j};s_{i,j}]\right),\qquad\widehat{S}_{i,j}=\prod_{t=1}^{j}p_{i,t},
\end{equation}
where $w$ is a learned weight vector (bias omitted) and $[\cdot;\cdot]$ concatenation.
Here $p_{i,j}$ estimates the probability of accepting token $j$ given its predecessors, and $\widehat{S}_{i,j}$ is its post-draft survival probability.
One linear layer suffices, as $h_{i,j}$ and $s_{i,j}$ already carry rich backbone and selector information.

\textbf{Budget distribution.}
The budget distributor selects the top-$B$ positions by $\widehat{S}_{i,j}$ across the batch, applying \dspark's \adaptveri~\cite{dspark} (\cref{sec:background}) to \thiswork's variable-length drafts.
The target verifies these $B$ tokens with the $N$ non-speculative ones as one packed batch.
As $\widehat{S}_{i,j}$ never increases in $j$, the selection forms a prefix of each draft.
This prefix is fixed before verification, which applies standard speculative sampling~\cite{leviathan23}, so the output distribution is unchanged.

\subsection{Training and Calibration}
\label{sec:training}

Both estimators are trained jointly with the \dflashp drafter by adding their losses to its objective, $\mathcal{L}=\mathcal{L}_{\mathrm{DFlash2}}+\lambda_{\mathrm{pre}}\mathcal{L}_{\mathrm{pre}}+\lambda_{\mathrm{post}}\mathcal{L}_{\mathrm{post}}$.
Instead of a binary label of whether each draft token was accepted, they learn from a \textit{soft acceptance target}, the probability that the token would be accepted, which gives a smoother training signal.
Following \dspark~\cite{dspark}, the target for position $j$ is its acceptance probability under speculative sampling, \(\alpha_j=\sum_{v\in\mathcal{C}_j}\min\{P_j(v),Q_j(v)\}\), with the target distribution \(P_j\) and the drafter's proposal \(Q_j\).
Unlike \dspark, the sum covers only \(\mathcal{C}_j\), the candidate tokens the path selector keeps, and \(Q_j\) is the proposal over them after the selector's choice.
For the pre-draft estimator, each training step samples a length \(d\), runs the backbone at exactly that length, and trains row \(z^{(d)}\) on that length's targets \(\alpha_{d,j}\) with binary cross-entropy, since bidirectional drafts of different lengths differ.
Gradients stop at \(c_i\) and \(e_i\), so this loss trains only the pre-draft estimator and leaves the drafter unchanged.
The post-draft estimator is trained against \(\alpha_j\) with a weighted binary cross-entropy that excludes the anchor.
Since each survival estimate multiplies the per-position probabilities, a small error at each position accumulates along the draft.
After training, we therefore recalibrate both estimators on Alpaca~\cite{alpaca} prompts, fitting a temperature for each position of the post-draft estimator, as in \dspark~\cite{dspark}, and fitting a scale and bias for each entry of the pre-draft estimator (see \cref{app:calibration}).

%% file: sections/5_implementation.tex
\section{Implementation}
\label{sec:implementation}

We implement \thiswork on vLLM~0.30.0~\cite{vllm} with about 4K lines of Python and Triton, and train the drafters and estimators with Speculators~\cite{speculators}.

\textbf{Scheduler optimization.}
\thiswork splits the scheduler so that the host never waits on the GPU: the host decides how many tokens to draft and verify, the GPU decides which ones, and the two exchange only scores and totals.
Before drafting in step \(t\), the host computes the totals of \cref{eq:joint}, \(\sum_i d_i\) and \(B\), from the pre-draft scores the GPU copied to it in step \(t-1\), and sends \(\sum_i d_i\) to the GPU.
A newly admitted request without scores on the host uses the others' mean scores (under 1\% of request-steps).
The GPU then runs the pre-draft estimator on the current target states, splits \(\sum_i d_i\) into per-request lengths \(d_i\) by ranking these fresh scores with a one-pass Triton kernel, and asynchronously copies the scores to the host for step \(t+1\).
Meanwhile, the host recomputes \(B\) with the verification step's actual non-speculative token count \(N\) and sizes that step at \(N+B\) tokens.
Using this \(B\), the GPU's budget distributor then picks the top-\(B\) positions by their post-draft estimates with the same score-ranking Triton kernel.

\textbf{Execution.}
Each decoding step replays captured CUDA graphs for the pre-draft estimator, the drafter, and the output stage, which applies the LM head, picks the draft tokens, and runs the post-draft estimator.
The drafting and verification latency tables are built by timing these graphs once.
Under tensor parallelism, small numerical differences across ranks could produce different schedules and stall collective operations.
\thiswork therefore broadcasts rank~0's pre-draft scores, draft tokens, post-draft estimates, and acceptance decisions at every step (each at most \(K(K+1)/2\) values per request) and uses rank~0's latency tables, so all ranks schedule from identical inputs.

\textbf{Hybrid and sparse-attention targets.}
The sparse-MLA kernels of the GLM models need each request's lengths on the host before launch.
For these targets, the host performs the per-request split and selects the verified positions itself, from the previous step's estimates, and the GPU executes them.
Stale estimates affect only which tokens are drafted and verified, not the output, as verification is lossless.

%% file: sections/6_evaluation.tex
\section{Evaluation}
\label{sec:evaluation}

\subsection{Experimental Setup}
\label{sec:eval:setup}

\textbf{Models.}
We use models that span dense, mixture-of-experts (MoE), and hybrid architectures with 14--744B parameters: \qwen~\cite{qwen3}, a 40-layer dense model served in BF16, \glmflash~\cite{glm53flash} (320B, 18B active), a hybrid MoE model whose 45 layers interleave KDA linear attention~\cite{kimilinear} and MLA with DSA sparse attention~\cite{deepseekv2,deepseekv32}, and \glm~\cite{glm5} (744B, 40B active), a 78-layer MoE model using MLA with DSA sparse attention.
Models are served with 8-way tensor parallelism (TP8), plus 8-way expert parallelism (EP8) for the MoE models, and both GLM models use NVFP4 routed experts and an FP8 KV cache.
We use the lowest reasoning effort and greedy decoding (temperature 0) unless noted.

\textbf{Hardware and software.}
We serve \qwen on eight NVIDIA H100 GPUs (96\,GB each), and train all drafters and serve \glmflash/\glm on eight NVIDIA B200 GPUs (180\,GB each), with vLLM~0.30.0, PyTorch~2.13, and CUDA~13.0 (host details in \cref{app:eval_workload_metric}).

\textbf{Baselines.}
We compare \thiswork, which jointly schedules drafting and verification, with autoregressive decoding without speculation (\textbf{\ar}) and three parallel-drafting baselines: \textbf{\dflashp}~\cite{dflash2}, a recent parallel drafter, which drafts and verifies a full block, \textbf{Batchwise (\dflashpp)}, \dflashp with vLLM's load-based schedule~\cite{vllmdynsd}, which caps the speculation length of the whole batch at a pre-calibrated value per load, and \textbf{\dspark}~\cite{dspark}, a goodput-aware parallel drafter with \adaptveri.
We also evaluate \textbf{\oursnojs}, \thiswork without its pre-draft estimator and joint scheduler (JS), which drafts full blocks and applies \adaptveri with \thiswork's post-draft estimator, isolating the gain of joint scheduling.
It thus applies the same \adaptveri as \dspark and differs mainly in the drafter backbone.
All parallel drafters follow their published training recipes (e.g., 5/6 layers) and use a maximum draft length of $K$=15, with $K$=7 for sensitivity (\cref{sec:eval:sensitivity}).
To show that WCS generalizes, we also apply it to \dspark and to two autoregressive drafters with $K$=5, \mtp~\cite{deepseekv3} and \eagle~\cite{eagle3} (\cref{sec:eval:other_drafters}).
Drafter training and the pre-deployment calibration details are in \cref{app:eval_drafter_training,app:calibration}.

\textbf{Workloads and metrics.}
Our main workload replays 20,000 multi-turn ShareGPT~\cite{sharegpt} conversations, and a decode-only variant releases each batch together after prefill and times 512 decode tokens per answer.
In closed-loop runs, we sweep the load, the number of active clients and thus the batch size, from one to the largest load each deployment fits (e.g., 512 on \qwen) and plot total output tokens per second (TPS) against \textit{user interactivity}, the per-user TPS given by the inverse of the median time per output token (TPOT, excluding the first token).
We compare methods at matched user interactivity as in \dspark~\cite{dspark}, interpolating between measured points and reporting only points where a method reaches at least half of its peak throughput.
In open-loop runs, requests arrive as a seeded Poisson process, and following \citet{alphaserve}, we report service level objective (SLO) attainment, the fraction of requests meeting both their time to first token (TTFT) and TPOT targets, and capacity, the arrival rate at which attainment falls to 90\% (\cref{app:eval_workload_metric}).

\subsection{Analysis on Serving Performance}
\label{sec:eval:serving}

\begin{figure}[t]
    \centering
    \includegraphics[width=\columnwidth]{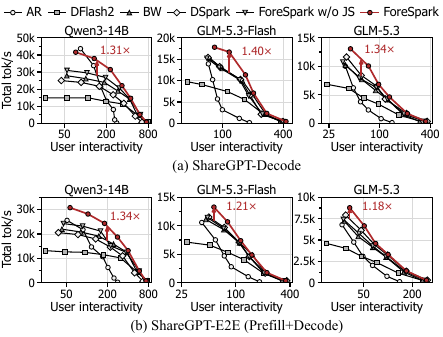}
    \caption{Total throughput versus user interactivity on (a) ShareGPT-Decode and (b) ShareGPT-E2E.
    Arrows mark the largest gain over the strongest baseline.}
    \label{fig:main_pareto}
\end{figure}

On the decode-only (ShareGPT-Decode) and end-to-end (ShareGPT-E2E) workloads of \cref{fig:main_pareto}, \thiswork achieves up to 1.31/1.40/1.34$\times$ and 1.34/1.21/1.18$\times$ the throughput of the strongest baseline at matched user interactivity (arrows) and 1.77/1.40/1.34$\times$ and 1.53/1.21/1.18$\times$ that of \dspark on \qwen/\glmflash/\glm.
The gains reach 1.37/1.46/1.46$\times$ and 1.43/1.24/1.30$\times$ over \oursnojs, which shares \thiswork's backbone, and 1.61/2.49/2.39$\times$ and 2.04/2.19/2.08$\times$ over \ar.

Adaptive verification lets \dspark and \oursnojs outperform \dflashp under high load, but both still draft every block in full and pay the drafter overhead.
\thiswork reduces this overhead with joint scheduling and outperforms both.
The gains of \thiswork over \dspark are larger on ShareGPT-Decode, since in ShareGPT-E2E prefill tokens share each step with decoding and add non-speculative tokens $N$ that draft scheduling does not reduce.
\dflashpp even outperforms \dspark at several loads, showing that its load-based cap is effective there.
However, it applies one length to the batch regardless of each request's acceptance, whereas \thiswork schedules drafts per request and outperforms it.

At a load (batch size) of one, where full-block drafting is preferable and lies within \thiswork's design space, \thiswork attains 1.15/1.07/0.96$\times$ and 1.11/0.98/0.98$\times$ the user interactivity of \dspark on ShareGPT-Decode and ShareGPT-E2E, with the small gap on \glm consistent with \dspark's slightly higher acceptance (\cref{sec:eval:qualitative}).
On \qwen ShareGPT-Decode, \ar has the highest throughput at the heaviest load, where verification is compute-bound, but \thiswork comes closest among speculative methods and surpasses \ar at higher user interactivity.
On \qwen ShareGPT-E2E, in contrast, \thiswork exceeds \ar even at the heaviest load.
We attribute this to long multi-turn chat histories in ShareGPT-E2E, which keep every method memory-bound, so extra verification stays computationally cheap and speculation remains worthwhile.

\begin{figure}[t]
    \centering
    \includegraphics[width=\columnwidth]{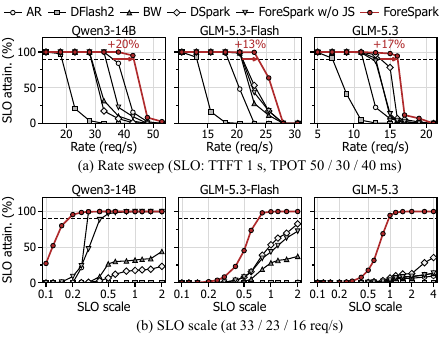}
\caption{SLO attainment on ShareGPT-E2E, following \citet{alphaserve}.
(a) Versus request rate, where arrows mark capacity gains at 90\% (dashed).
(b) At a fixed rate, with both thresholds scaled.}
    \label{fig:main_slo}
\end{figure}

\textbf{SLO attainment.}
Open-loop runs use the end-to-end workload, as a decode-only run cannot measure TTFT.
At 90\% attainment (\cref{fig:main_slo}(a)), \thiswork sustains 1.20/1.13/1.17$\times$ the request rate of the strongest baseline, 1.51/1.13/1.17$\times$ that of \dspark, and 1.29/1.13/1.28$\times$ that of \oursnojs on \qwen/\glmflash/\glm.
Once requests queue, extra step time delays active generations and queued prefills, so reducing unproductive drafting and verification keeps both SLOs met at higher rates.

\cref{fig:main_slo}(b) fixes the request rate just above \dspark's capacity (33/23/16 requests/s) and scales the TTFT and TPOT thresholds by a common factor.
At the original thresholds, \thiswork meets the SLOs for 100/100/95\% of requests, against 17/53/7\% for \dspark and 100/43/4\% for \oursnojs.
\thiswork keeps attainment above 90\% even when the thresholds are tightened to 0.2/0.8/1.0$\times$, whereas \dspark stays below 90\% even when they are relaxed to 2$\times$ on \qwen and \glmflash and 4$\times$ on \glm.
At this rate, \dspark's queue keeps growing, so looser thresholds recover only a fraction of its requests, while \thiswork still has headroom.

\subsection{Execution Time Breakdown Analysis}
\label{sec:eval:breakdown}
\begin{figure}[t]
    \centering
    \includegraphics[width=\columnwidth]{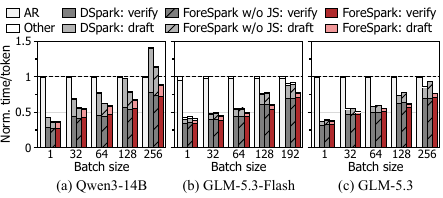}
    \caption{Time per output token on ShareGPT-Decode, normalized to \ar at each batch size.}
    \label{fig:breakdown}
\end{figure}

\cref{fig:breakdown} breaks down the time per output token.
At the largest batch size, \thiswork reduces drafting time by 76/68/67\% relative to \dspark on \qwen/\glmflash/\glm, as it packs only the scheduled draft positions (\cref{sec:postdraft}), and by 62/70/76\% even against \oursnojs, which shares the same backbone, so the saving comes from joint scheduling rather than the backbone.
Joint scheduling cuts drafts from 15 to 2.04/4.75/3.55 tokens per request at the largest batch size, and \thiswork verifies nearly all of them (\cref{app:token_breakdown}).
Total time drops by 37/13/11\% relative to \dspark, and on \qwen, \dspark and \oursnojs need 1.41$\times$ and 1.14$\times$ \ar's time per token, whereas \thiswork needs 0.89$\times$.
Verification time also drops at some batch sizes (e.g., 12\% at 128 requests on \glmflash) despite the same post-draft estimator as \oursnojs, which we attribute to joint scheduling sizing drafts and verification together.

\subsection{Analysis on Joint Scheduler}
\label{sec:eval:ablation}

\begin{figure}[t]
    \centering
    \includegraphics[width=\columnwidth]{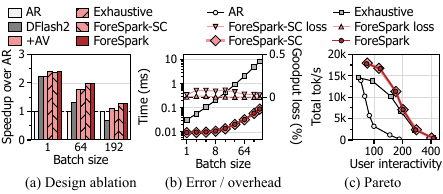}
    \caption{Scheduler analysis on \glmflash.
    (a) Decode-only speedup over \ar for design variants.
    (b) Comparison with exhaustive search.
    (c) Total throughput versus user interactivity.}
    \label{fig:ablation_scheduler}
\end{figure}

On \glmflash with a common \dflashp backbone (\cref{fig:ablation_scheduler}(a)), adaptive verification alone (\texttt{+AV}, i.e., \oursnojs) improves throughput by up to 1.56$\times$ over full verification (\dflashp), and joint scheduling adds up to 1.17$\times$ on top.
This gain relies on a computationally cheap search, as an exhaustive scheduler with the same inputs (\texttt{Exhaustive}) falls below \texttt{+AV} at batch sizes of 64 and 192 and trails \thiswork by up to 1.22$\times$.
\thiswork decides a schedule in 88\,$\mu$s at 192 requests, against 8.2\,ms for \texttt{Exhaustive}, used only as an optimality reference, with an average goodput loss of at most 0.01\% (\cref{fig:ablation_scheduler}(b)).

\thiswork-SC (single curve) replaces the per-length estimator with one shared length-$K$ curve, which assumes prefix structure and makes each $R_i$ concave.
Despite the token differences in \cref{sec:motivation}, it predicts acceptance as accurately and matches \thiswork in all panels, so the backbone admits this simpler form, which \thiswork forgoes for generality.

\subsection{Sensitivity Studies}
\label{sec:eval:sensitivity}

\textbf{Dataset.}
\thiswork's advantage carries over to math and code workloads, with up to 1.49/1.38/1.24$\times$ the throughput of \dspark on \qwen/\glmflash/\glm (\cref{app:dataset}), as it sets each request's draft length from its current context rather than from a workload-specific setting.

\begin{figure}[t]
    \centering
    \includegraphics[width=\columnwidth]{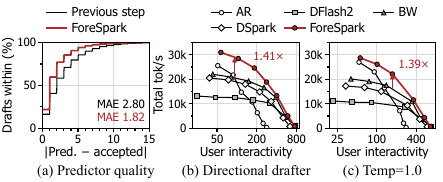}
    \caption{(a) Share of drafts within a given acceptance-length error.
    (b) Total throughput versus user interactivity with a directional drafter.
    (c) Total throughput versus user interactivity of \qwen on ShareGPT-E2E with temperature 1.0.}
    \label{fig:pred_causal}
\end{figure}

\textbf{Predictor quality.}
Although the previous step's acceptance is infeasible under adaptive verification (\cref{sec:motivation}), we compare against it over 1.48M full-block drafts of \qwen ($K$=15, no adaptive verification), where \thiswork predicts acceptance within one token for 60\% of drafts versus 41\%, and lowers the mean absolute error from 2.80 to 1.82 tokens (\cref{fig:pred_causal}(a)), so the target model's current features forecast the upcoming draft better than historical acceptance.

\textbf{Drafter type.}
With a directional drafter, i.e., causal attention within the block (\cref{fig:pred_causal}(b)), \thiswork still improves throughput over \dspark by up to 1.55$\times$ at matched user interactivity.
Full-block drafting remains overhead under load even for this drafter, which joint scheduling reduces.

\textbf{Temperature.}
With temperature 1.0 on \qwen ShareGPT-E2E (\cref{fig:pred_causal}(c)), \thiswork achieves up to 1.39$\times$ the throughput of the strongest baseline (\dflashpp), 1.76$\times$ that of \dspark, and 1.70$\times$ that of \ar, compared with 1.34/1.53/2.04$\times$ under greedy decoding.
Sampling lowers acceptance, which narrows the margin over \ar but widens the gap to the speculative baselines, as full-block drafting wastes more work when fewer drafts are accepted.

\textbf{Draft length.}
With $K$=7, \thiswork achieves up to 1.32$\times$ the throughput of \dspark on \qwen and up to 1.24/1.16$\times$ that of the strongest baseline on \glmflash/\glm, so its gains hold although a smaller $K$ leaves less drafter work to avoid.
A larger $K$ gives \thiswork more choices while joint scheduling limits the extra drafting, easing the barrier noted in \cref{sec:motivation}.
On math and code with \qwen, $K$=15 helps at low load, but at high load it lowers every speculative baseline's throughput below its $K$=7 version.
\thiswork keeps the low-load gain and still achieves the highest throughput at high load (\cref{app:k_7}).

\subsection{Analysis on Acceptance Length and Overhead}
\label{sec:eval:accept_lengt}
\label{sec:eval:qualitative}

\cref{fig:accept_train} (\cref{app:drafter_cost}) tests whether \thiswork's gains stem from a stronger or more expensive drafter.
At $K$=15 without adaptive verification, \thiswork reaches mean acceptance lengths of 4.38/3.33/3.37 on \qwen/\glmflash/\glm over SPEED-Bench's 11 categories~\cite{speedbench}, within 3.2\% of \dspark and equal to \dflashp since its estimators only score drafts from the same backbone, so the serving gains do not rely on higher acceptance.
Training costs little extra, taking 1.12/1.05/1.11$\times$ the time per iteration of \dspark and only 3--5\% more than \dflashp, since length sampling trains one pre-draft row per step instead of running the drafter at every length (\cref{sec:training}).
The pre-/post-draft estimators add only 0.1--0.2\% parameters, and the drafter is 3.7--5.0\% larger than \dspark's due to the \dflashp backbone.

\subsection{Analysis on WCS with Other Drafters}
\label{sec:eval:other_drafters}
\label{sec:eval:autoreg}

\begin{figure}[t]
    \centering
    \includegraphics[width=\columnwidth]{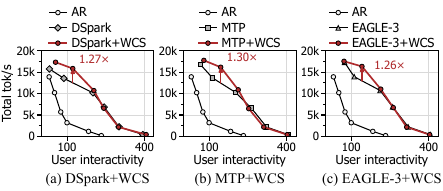}
    \caption{Total throughput versus user interactivity of other drafters with and without WCS on \glmflash.}
    \label{fig:wcs_drafters}
\end{figure}

Since WCS needs only a drafter latency profile and a pre-draft estimator on the target's context feature and anchor embedding (\cref{sec:predraft}), it is not tied to \thiswork's drafter, and \cref{fig:wcs_drafters} applies it to the parallel drafter \dspark and the autoregressive drafters \mtp and \eagle, keeping each drafter's maximum draft length and serving the same checkpoint with and without WCS.
At matched user interactivity, WCS improves throughput by up to 1.27/1.30/1.26$\times$ over \dspark, \mtp, and \eagle (see \cref{app:orthogonal} for implementation details).

%% file: sections/7_discussion.tex
\section{Discussion}

\textbf{Where \thiswork helps most.}
Since full-block drafting is one of its schedules, \thiswork stays comparable to \dspark at low load while extending speculative decoding to higher loads and decode-dominated workloads, where speculation is usually considered unsuitable (\cref{sec:eval:serving,sec:eval:breakdown}).

\textbf{Remaining headroom.}
Against an oracle that knows each draft's acceptance, both pre-/post-draft estimators remain conservative.
This partly stems from our calibration on the external Alpaca dataset, which assumes an unknown workload.
Further gains thus lie in workload-matched calibration or better workload-adaptive estimators.

\textbf{Fairness.}
Global sorting may leave a request undrafted in a step, which never changes outputs but can skew per-request speedups.
Per-request minimum budgets could fix this.

%% file: sections/8_related_work.tex
\section{Related Work}
\label{sec:related}

\textbf{Goodput-aware speculation in LLM serving.}
Complementary to drafter designs (\cref{app:related}), serving systems schedule prefill and decoding for throughput or goodput~\cite{orca,vllm,sglang,sarathiserve,distserve}, but under load extra verification competes with other requests and can reduce or even reverse the benefit of speculation~\cite{specbatch,turbospec,specillusion}.
To address this, TurboSpec~\cite{turbospec} and Nightjar~\cite{nightjar} choose one speculation length per batch, and AdaServe~\cite{adaserve}, AdaSpec~\cite{adaspec}, FASER~\cite{faser}, and ECHO~\cite{echo} size it per request or tree, often under SLO or budget constraints, mostly for autoregressive or tree drafters.
Verification budgeting selects the most promising draft tokens across requests~\cite{tetris}, including for parallel drafters~\cite{dspark,dcut,angelspec,dscale,adaflash}, but only after a full block is drafted, leaving the drafter overhead.

\textbf{Draft length control.}
Adaptive draft lengths are mostly studied for autoregressive drafters through learned stopping predictors~\cite{specdecpp,disco}, single-request length regression~\cite{adaeagle}, drafting overlapped with verification~\cite{pearl}, tree search~\cite{opttree}, or online tuning~\cite{banditspec}.
For parallel drafters, DiffuSpec~\cite{diffuspec} sizes drafts from recent acceptance, FailFast~\cite{failfast} and DLoop~\cite{dloop} keep drafting while confident at the cost of extra passes, and TreeSpark~\cite{treespark} prunes a draft tree by path survival and shrinks it under load but builds it from a full backbone pass.
None of them sizes each request's draft before a parallel drafter's pass from current-context acceptance estimates, jointly with a verification budget.

%% file: sections/9_conclusion.tex
\section{Conclusion}
\label{sec:conclusion}

We presented WCS and its prototype \thiswork, which estimate acceptance before drafting, jointly allocate per-request draft lengths and the verification budget, and spend it with post-draft confidence, improving throughput by up to 1.77$\times$ over \dspark.
We hope WCS guides how future parallel drafters are designed and scheduled for LLM serving.

%% file: sections/etc_ai_usage_disclosure.tex
\section*{AI Usage Disclosure}
We used Anthropic's Claude models (e.g., Opus 5.5) as an assistant in parts of this work.
All research ideas, design decisions, and conclusions are the authors' own, and the authors reviewed and verified every AI-assisted output.

\textbf{Ideation and motivation.}
The research problem, the motivational observations, and the design of work-conserving speculation are the authors' own, and the motivational measurements were designed and run by the authors.

\textbf{Literature search.}
We used the assistant to search for related work and to check whether our idea had been proposed before.
The authors read the identified papers themselves, and every citation was verified against its primary source.

\textbf{Code.}
The authors implemented \thiswork with AI assistance for code generation, most notably to optimize the Triton kernel for variable-length drafting.
All AI-assisted code was reviewed and tested by the authors.

\textbf{Experiment design and data.}
The authors designed and ran all experiments, using the assistant for feedback in discussion and to write plotting scripts.
No benchmark or evaluation data were generated with AI tools.

\textbf{Writing.}
The authors wrote the content of the paper.
Under the authors' direction, the assistant was used to revise drafts for clarity and concision and to check consistency between the text, figures, and notation.
The authors checked and substantially revised all AI-assisted text and take full responsibility for the content.

%% file: sections/etc_appendix.tex
\section{Drafter Training Details}
\label{app:eval_drafter_training}

\textbf{Training.}
We train all drafters with Speculators~\cite{speculators} on 958K user turns from Magpie~\cite{magpie} and UltraChat~\cite{ultrachat}, after removing the SPEED-Bench~\cite{speedbench} prompts.
To generate training data, the target model regenerates every response with its recommended sampling settings, up to 8,192 tokens, with thinking disabled on \qwen and the lowest reasoning effort on the GLM models.
During training, a vLLM server running the target supplies the hidden states of the captured layers (\cref{tab:setup}).
A trainer trains the parallel drafters on the regenerated data, using \(K\)=15 with bidirectional attention within the block and 8,192-token sequences with up to 3,072 anchors.
Their loss is 0.1 times cross-entropy plus 0.9 times total variation, with position weights that decay exponentially.
We train for three epochs with AdamW and a cosine schedule at the learning rates of the published recipes (\(6\times10^{-4}\) for \dflashp and \(3\times10^{-4}\) for \dspark).
\dflashp, \dflashpp, and \oursnojs use \thiswork's backbone, so \dflashp drafts the same tokens as \thiswork at length \(K\).
For other drafter hyperparameters, such as the number of layers and the captured target layers, we use the settings of widely adopted published drafters.

\textbf{Pre-draft labels.}
The training length \(d\) of \cref{sec:training} cycles through \(K,1,\ldots,K-1\) across steps.
When \(d<K\), the drafter first re-drafts at length \(d\) from every fourth anchor without gradients, and the labels \(\alpha_{d,j}\) compare only these drafted positions with the target distributions of the step.

\textbf{Other drafters.}
The directional drafters of \cref{fig:pred_causal}(b) are retrained with causal attention within the block.
\eagle is a one-layer drafter trained with the Speculators recipe for three epochs, and \mtp is \glmflash's own multi-token-prediction layer (\cref{app:orthogonal}).

\begin{table*}[t]
    \centering
    \caption{Training and serving configuration, identical for all methods on a model. All drafters train on eight B200 GPUs, and the \glmflash drafters train against its FP8 release. Loads take the values 1, 8, 32, 64, 128, 256, and 512 within each range.}
    \label{tab:setup}
    \setlength{\tabcolsep}{30pt}
    \resizebox{\textwidth}{!}{%
    \begin{tabular}{lccc}
    \toprule
     & \qwen & \glmflash & \glm \\
    \midrule
    Captured target layers & 1, 10, 19, 28, 37 & 6, 15, 25, 34, 43 & 6, 20, 34, 48, 62, 76 \\
    Drafter layers \(\times\) hidden size & \(5\times5{,}120\) & \(5\times4{,}096\) & \(6\times6{,}144\) \\
    Training GPUs (server / trainer), steps & 2 / 6, 112K & 4 / 4, 154K & 4 / 4, 162K \\
    \midrule
    Serving GPUs and parallelism & 8\(\times\)H100, TP8 & 8\(\times\)B200, TP8 + EP8 & 8\(\times\)B200, TP8 + EP8 \\
    Weights / KV cache & BF16 / BF16 & NVFP4 experts / FP8 & NVFP4 experts / FP8 \\
    Max.\ requests / tokens per step & 512 / 24,576 & 256 / 24,576 & 256 / 24,576 \\
    CUDA graphs up to (tokens) & 8,192 & 4,096 & 4,096 \\
    Closed-loop / decode-only loads & 1--512 / 1--512 & 1--256 / 1--192 & 1--256 / 1--256 \\
    Request rates (req/s) & 4--53 & 4--30.5 & 2.5--21 \\
    SLO (TTFT / TPOT) & 1\,s / 50\,ms & 1\,s / 30\,ms & 1\,s / 40\,ms \\
    \bottomrule
    \end{tabular}}
\end{table*}

\section{Calibration Details}
\label{app:calibration}
After training, we calibrate both pre-/post-draft estimators once per model with their weights fixed, using up to 4,000 Alpaca prompts~\cite{alpaca} that appear in neither the training data nor any benchmark.
The calibration runs use greedy decoding and verify every drafted position, so every position has an acceptance label.

\textbf{Post-draft estimator.}
Let \(\ell_{i,j}=w^{\top}[h_{i,j};s_{i,j}]\) be the logit of the post-draft estimator (\cref{sec:postdraft}).
Following \dspark~\cite{dspark}, each position \(j\) gets a temperature \(\theta_j\), so that \(p_{i,j}=\sigma(\ell_{i,j}/\theta_j)\) and \(\widehat{S}_{i,j}=\prod_{t\leq j}p_{i,t}\).
The temperatures are fitted in order from \(j\)=1 to \(K\), as
\begin{equation}
\theta_j=\operatorname*{arg\,min}_{\theta\in\Theta}\operatorname{ECE}\bigl(\widehat{S}_{i,j}(\theta),\,\mathbf{1}[A_{i,K}\geq j]\bigr),
\label{eq:sts}
\end{equation}
where \(\widehat{S}_{i,j}(\theta)\) uses \(\theta\) at position \(j\) and the already fitted temperatures before it, \(A_{i,K}\) is the number of accepted tokens when request \(i\) drafts the full length \(K\) (\cref{sec:predraft}), ECE is the expected calibration error over 15 equal-width bins, and the grid \(\Theta\) holds 41 values evenly spaced in \(\log\theta\) from 0.1 to 10.
The traces come from 1,000 of the prompts, drafted at the full length \(K\) with up to 256 output tokens each at a load of 16.
\dspark is calibrated in the same way, and the temperature-1.0 runs reuse the greedy fit.

\textbf{Pre-draft estimator.}
Since each draft length yields a distinct proposal, we calibrate each entry \((d,j)\) of the triangle in \cref{sec:predraft} on a run at exactly length \(d\).
The calibrated conditional acceptance is \(\sigma(a_{d,j}z^{(d)}_{i,j}+b_{d,j})\), where the scale \(a_{d,j}\) and bias \(b_{d,j}\) come from a logistic regression of whether position \(j\) is accepted, fitted on the drafts whose first \(j-1\) tokens are accepted.
One run per length \(d=1,\ldots,15\) on the 4,000 prompts, sweeping the load up to 512 requests (256 on \glm), yields all \(K(K+1)/2\) pairs.
A directional drafter needs only the length-\(K\) run, since its shorter drafts are prefixes of the longer one.

\textbf{Latency tables and \dflashpp.}
\thiswork takes drafting latencies from its startup profile (\cref{sec:implementation}) and verification latencies from offline profiles.
Each verification entry is the median target forward time for a given batch size and number of verified tokens per request, over at least 30 decode-only steps.
For the length cap of \dflashpp in vLLM~0.30, we run every cap \(k=1,\ldots,15\) at every load on the 4,000 prompts and assign each load range its fastest cap.

\section{Evaluation Workload and Metric Details}
\label{app:eval_workload_metric}

The \qwen server has two Intel Xeon Platinum 8480CL CPUs and the B200 servers have two AMD EPYC 9654 CPUs, each with 2\,TB of DDR5 DRAM.
Each method runs alone on the node with the settings of \cref{tab:setup}.

\textbf{ShareGPT-E2E.}
We take the first 20,000 ShareGPT conversations with at least four turns and form a stream of 88,051 requests, one per user turn.
Each request carries the conversation so far, up to about 16K tokens, and consecutive turns of a conversation are about 512 requests apart.
A request ends at the end-of-turn token or after 4,096 output tokens.
Each closed-loop point starts from an empty prefix cache and warms up on held-out conversations for at least 30\,s, until the batch reaches 90\% of the load or stops growing.
It then measures at least 128 requests over at least 120\,s, and every method replays the same request sequence.
We compare methods at matched user interactivity as in \dspark~\cite{dspark}.
We interpolate each curve linearly in log user interactivity between adjacent loads and report ratios only where both methods run at no less than half of their peak throughput.

\textbf{ShareGPT-Decode.}
Each batch of \(C\) requests, where \(C\) is the batch size, is prefilled from an empty prefix cache and held at an engine-side barrier until every request has its first token, and then all requests are released together.
Each request is timed over the 512 tokens after its first token, cut at its first end-of-turn token (e.g., \texttt{<|im\_end|>} on \qwen).
To keep such cuts rare, we first decode the 4,096 ShareGPT prompts closest to 1K characters greedily with \ar and keep only those whose answer reaches 641 tokens (1,572/618/646 prompts on \qwen/\glmflash/\glm).
Since verification is lossless, every method then times the same answers, and only 2\% of the timed answers on \qwen still end early, because numerical differences can change the greedy choice between nearly tied tokens.
A batch's throughput is the sum of its requests' token rates over their timed spans, and its user interactivity is the median of these rates.

\textbf{Open-loop tests for SLO experiments.}
Requests arrive as a seeded Poisson process that is identical for every method.
At rate \(r\), \(30r\) warm-up arrivals precede \(120r\) measured ones, and requests that fail or do not finish count as SLO misses.
All methods run at the same request rates, chosen from short pilot runs.
The sweep uses at most 12 rates: two light-load rates, then evenly spaced rates up to just past the point where \thiswork, the method with the highest capacity, meets the SLOs for fewer than half of the requests.

\textbf{Breakdown and acceptance.}
\cref{fig:breakdown} normalizes each method's time per output token to \ar, i.e., \ar's throughput divided by the method's.
It splits it into verification, drafting, and other time in proportion to step times from vLLM's CUDA-event step timer, which costs about 0.03\% of wall time.
We measure acceptance length, one plus the accepted draft tokens per verification step, on SPEED-Bench's 880 qualitative prompts~\cite{speedbench} at a load of 32, verifying every drafted token at \(K\)=15.

\section{Number of Drafted and Verified Tokens with Joint Scheduling}
\label{app:token_breakdown}

\begin{figure}[t]
    \centering
    \includegraphics[width=\columnwidth]{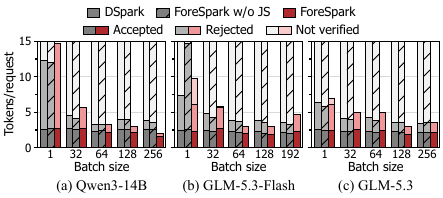}
    \caption{Average draft tokens per request per step on ShareGPT-Decode ($K$=15), split into accepted, rejected, and not verified tokens, for the runs of \cref{fig:breakdown}.}
    \label{fig:token_breakdown}
\end{figure}

\cref{fig:token_breakdown} counts how many tokens each method drafts, verifies, and accepts per request in each step of \cref{fig:breakdown}, showing how the time breakdown follows the scheduled lengths.
\dspark and \oursnojs always draft the full block of 15 tokens, but from 32 requests on, adaptive verification keeps only 3.26--4.77 of them and discards the remaining 68--78\% of the drafter's output unverified.
\thiswork drafts only 2.04--5.76 tokens per request at these batch sizes and verifies 99.7--100\% of them, which explains its lower drafting time in \cref{sec:eval:breakdown}.
\thiswork verifies between 1.75 fewer and 1.21 more tokens per request than \dspark, since joint scheduling balances drafting and verification rather than shrinking both.
At a load of one, \thiswork drafts 14.71/9.72/7.00 tokens per request, and on \glmflash and \glm, it over-drafts, leaving 3.67/0.87 for the post-draft stage to drop (\cref{sec:scheduler}).

\section{Dataset Sensitivity: Math and Code Workload}
\label{app:dataset}

\begin{figure}[t]
    \centering
    \includegraphics[width=\columnwidth]{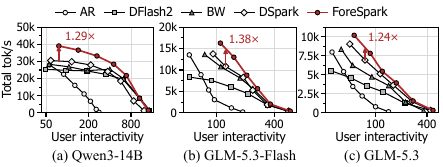}
    \caption{Total throughput versus user interactivity on math and code.
    Arrows mark the largest gain over the strongest baseline.}
    \label{fig:mathcode}
\end{figure}

To show that \thiswork's advantages are not tied to one workload, we also evaluate it on a math and code workload in \cref{fig:mathcode}.
The workload draws 2,855 single-turn prompts from GSM8K~\cite{gsm8k}, MATH-500~\cite{mathdataset,lightman}, AIME25~\cite{aime25}, HumanEval~\cite{humaneval}, MBPP~\cite{mbpp}, and LiveCodeBench~\cite{livecodebench} v5--v6, each with its set's usual instruction.
It runs in closed loop like ShareGPT-E2E, with the prefix cache disabled.
At matched user interactivity, \thiswork reaches up to 1.49/1.38/1.24\(\times\) the throughput of \dspark and up to 1.29/1.38/1.24\(\times\) that of the strongest baseline on \qwen/\glmflash/\glm (arrows in \cref{fig:mathcode}).
These answers are more predictable than ShareGPT's (\cref{tab:mathcode_accept}), so \thiswork outperforms \ar at every load, unlike on \qwen ShareGPT-Decode (\cref{sec:eval:serving}).
As on ShareGPT, the gains appear at high load, where \thiswork shortens drafts of requests with low expected acceptance and keeps long drafts for the rest.
The curves converge at high user interactivity, where full-block drafting is preferred and within \thiswork's design space.

\section{Acceptance Length / Drafter Cost}
\label{app:drafter_cost}

To show that the serving gains of \thiswork come from joint scheduling rather than from a stronger or more expensive drafter, we analyze the acceptance lengths of the drafters in \cref{tab:speed_accept}.
We use SPEED-Bench's qualitative split, which covers 11 categories~\cite{speedbench}.
\thiswork reaches average acceptance lengths of 4.38/3.33/3.37 on \qwen/\glmflash/\glm, against 4.28/3.35/3.48 for \dspark, staying within 3.2\% of \dspark.
\thiswork and \dflashp share one backbone and have identical acceptance, because the pre-/post-draft estimators only score drafts and never change the drafted tokens.
The serving gains over \dflashp therefore come from how \thiswork allocates drafting and verification, and those over \dspark do not rely on a higher token yield per full draft.

\begin{table}[t]
    \centering
    \caption{Acceptance length by SPEED-Bench category at \(K\)=15 without adaptive verification (80 prompts per category at a load of 32). \dflashp shares \thiswork's drafter.}
    \label{tab:speed_accept}
    \setlength{\tabcolsep}{5pt}
    \resizebox{\columnwidth}{!}{%
    \begin{tabular}{@{}lcccccc@{}}
    \toprule
     & \multicolumn{2}{c}{\qwen} & \multicolumn{2}{c}{\glmflash} & \multicolumn{2}{c}{\glm} \\
    \cmidrule(lr){2-3} \cmidrule(lr){4-5} \cmidrule(lr){6-7}
    Category & \dspark & \thiswork & \dspark & \thiswork & \dspark & \thiswork \\
    \midrule
    Coding & 5.918 & 5.863 & 3.753 & 3.855 & 4.038 & 3.979 \\
    Humanities & 4.495 & 4.715 & 2.947 & 2.950 & 3.101 & 3.021 \\
    Math & 6.211 & 6.556 & 3.185 & 3.154 & 3.291 & 3.425 \\
    Multilingual & 2.929 & 3.115 & 3.424 & 3.438 & 3.201 & 3.232 \\
    QA & 3.335 & 3.277 & 3.027 & 2.889 & 3.336 & 2.929 \\
    RAG & 4.378 & 4.445 & 4.662 & 4.838 & 4.775 & 4.752 \\
    Reasoning & 5.146 & 5.362 & 3.912 & 3.763 & 4.214 & 3.679 \\
    Roleplay & 2.539 & 2.532 & 2.222 & 2.145 & 2.257 & 2.229 \\
    STEM & 5.153 & 5.299 & 2.930 & 2.996 & 3.184 & 3.069 \\
    Summarization & 3.477 & 3.512 & 4.050 & 3.992 & 4.100 & 4.056 \\
    Writing & 3.476 & 3.488 & 2.704 & 2.612 & 2.737 & 2.654 \\
    \midrule
    Average & 4.278 & 4.379 & 3.347 & 3.330 & 3.476 & 3.366 \\
    \bottomrule
    \end{tabular}}
\end{table}

\begin{figure}[t]
    \centering
    \includegraphics[width=\columnwidth]{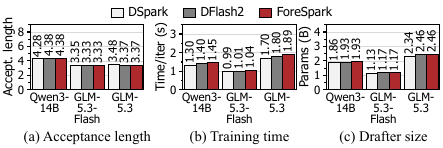}
    \caption{Drafter quality and cost.
    (a) Acceptance length on SPEED-Bench ($K$=15, no adaptive verification, per category in \cref{tab:speed_accept}).
    (b) Training time per iteration.
    (c) Parameters, excluding the embedding and LM head shared with the target.}
    \label{fig:accept_train}
\end{figure}

\cref{fig:accept_train}(b) reports the training time per iteration, measured in fresh runs at $K$=15 with the GPU splits of \cref{tab:setup} and averaged over minutes 10--15 after the first step.
\thiswork takes 1.45/1.04/1.89\,s per iteration, which is 1.12/1.05/1.11$\times$ that of \dspark (1.30/0.99/1.70\,s) and only 3--5\% more than \dflashp (1.40/1.01/1.80\,s).
This overhead stays small because length sampling trains only one row of the pre-draft estimator per step, rather than running the drafter for every candidate length (\cref{sec:training}).
With the training steps of \cref{tab:setup} (112K/154K/162K over three epochs), \thiswork's training takes about 45/44/85 hours on eight B200 GPUs, against 44/43/81 hours for \dflashp.

\cref{fig:accept_train}(c) reports drafter parameters, excluding the embedding and LM head that every drafter reuses from the target.
\thiswork has 1.93/1.17/2.46B parameters, almost the same as \dflashp, since the two estimators add only 0.1--0.2\%, and 3.7--5.0\% more than \dspark (1.86/1.13/2.34B) because of the \dflashp backbone.

\section{Discussion on Draft Length (\texorpdfstring{$K$}{K})}
\label{app:k_7}

\begin{figure}[t]
    \centering
    \includegraphics[width=\columnwidth]{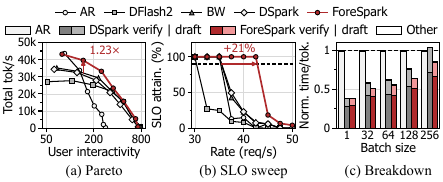}
    \caption{\qwen with \(K\)=7: (a) total throughput versus user interactivity on ShareGPT-Decode, (b) SLO attainment versus request rate on ShareGPT-E2E, and (c) time per output token on ShareGPT-Decode, normalized to \ar at each batch size.}
    \label{fig:k7}
\end{figure}

\begin{figure}[t]
    \centering
    \includegraphics[width=\columnwidth]{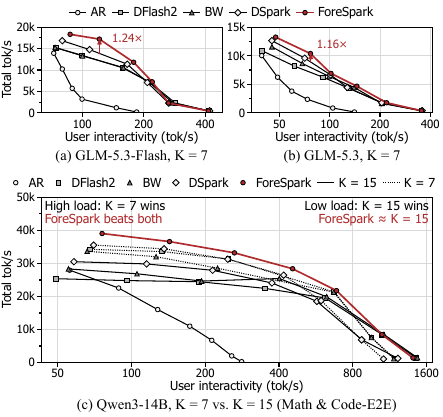}
    \caption{ShareGPT-Decode with $K$=7 on (a) \glmflash and (b) \glm, where arrows mark the largest gain over the strongest baseline.
    (c) A larger $K$ widens \thiswork's choice of draft lengths, which yields a better Pareto frontier of total throughput versus user interactivity (\qwen on the math and code workload, with $K$=7 dotted and $K$=15 solid).}
    \label{fig:k7_analysis}
\end{figure}

\textbf{Lower $K$ ($K$=7) sensitivity.}
To demonstrate that \thiswork still benefits drafters with a smaller \(K\), we train the drafters with \(K\)=7 using the same recipe.
\cref{fig:k7} shows that \thiswork achieves up to 1.23\(\times\) the decode-only throughput of the strongest baseline, \dflashpp, and 1.32\(\times\) that of \dspark at matched user interactivity (\cref{fig:k7}(a)).
It also sustains 1.21\(\times\) the request rate of \dspark at 90\% SLO attainment (42.8 versus 35.5 requests/s, \cref{fig:k7}(b)), and at 256 requests it spends 44\% less drafting time per output token than \dspark (\cref{fig:k7}(c)).
On larger models, \thiswork achieves up to 1.24\(\times\) and 1.16\(\times\) the throughput of the strongest baseline on \glmflash and \glm, as shown in \cref{fig:k7_analysis}(a) and (b).
A smaller \(K\) leaves less drafter work to avoid, yet the scheduler still shortens drafts whose expected yield does not justify their cost.

\begin{table}[t]
    \centering
    \caption{Acceptance length by math and code set at \(K\)=7 and \(K\)=15 without adaptive verification (80 prompts per set and all 30 of AIME25, at a load of 32), at most \(K+1\). \dflashp shares \thiswork's drafter.}
    \label{tab:mathcode_accept}
    \setlength{\tabcolsep}{1pt}
    \resizebox{\columnwidth}{!}{%
    \begin{tabular}{@{}lcccccccccccc@{}}
    \toprule
     & \multicolumn{4}{c}{\qwen} & \multicolumn{4}{c}{\glmflash} & \multicolumn{4}{c}{\glm} \\
    \cmidrule(lr){2-5} \cmidrule(lr){6-9} \cmidrule(lr){10-13}
     & \multicolumn{2}{c}{\dspark} & \multicolumn{2}{c}{\thiswork} & \multicolumn{2}{c}{\dspark} & \multicolumn{2}{c}{\thiswork} & \multicolumn{2}{c}{\dspark} & \multicolumn{2}{c}{\thiswork} \\
    \cmidrule(lr){2-3} \cmidrule(lr){4-5} \cmidrule(lr){6-7} \cmidrule(lr){8-9} \cmidrule(lr){10-11} \cmidrule(lr){12-13}
    Set & \(K\)=7 & \(K\)=15 & \(K\)=7 & \(K\)=15 & \(K\)=7 & \(K\)=15 & \(K\)=7 & \(K\)=15 & \(K\)=7 & \(K\)=15 & \(K\)=7 & \(K\)=15 \\
    \midrule
    GSM8K & 5.85 & 7.47 & 5.94 & 7.93 & 4.85 & 5.35 & 4.84 & 5.53 & 4.67 & 5.12 & 4.82 & 5.27 \\
    MATH-500 & 6.21 & 8.04 & 6.32 & 8.75 & 4.50 & 5.29 & 4.49 & 5.13 & 4.58 & 4.90 & 4.56 & 5.30 \\
    AIME25 & 5.77 & 7.66 & 6.00 & 8.07 & 3.46 & 3.68 & 3.53 & 3.67 & 3.73 & 3.93 & 3.75 & 3.73 \\
    HumanEval & 5.70 & 7.37 & 5.98 & 8.04 & 6.76 & 9.96 & 6.82 & 10.26 & 6.00 & 7.76 & 5.32 & 8.00 \\
    MBPP & 5.19 & 6.02 & 5.30 & 6.25 & 3.91 & 4.30 & 4.28 & 4.41 & 4.14 & 4.70 & 4.56 & 4.97 \\
    LiveCodeBench & 5.46 & 6.75 & 5.53 & 6.85 & 2.90 & 2.89 & 2.90 & 2.84 & 3.11 & 3.28 & 3.04 & 3.25 \\
    \midrule
    Average & 5.70 & 7.22 & 5.85 & 7.65 & 4.40 & 5.24 & 4.48 & 5.31 & 4.37 & 4.95 & 4.34 & 5.09 \\
    \(K\)=15 / \(K\)=7 & \multicolumn{2}{c}{1.27\(\times\)} & \multicolumn{2}{c}{1.31\(\times\)} & \multicolumn{2}{c}{1.19\(\times\)} & \multicolumn{2}{c}{1.19\(\times\)} & \multicolumn{2}{c}{1.13\(\times\)} & \multicolumn{2}{c}{1.17\(\times\)} \\
    \bottomrule
    \end{tabular}}
\end{table}

\textbf{Drafter overhead limits the use of a larger $K$.}
In actual deployment, a larger \(K\) raises acceptance length, but the extra drafter overhead limits its use in LLM serving.
\cref{tab:mathcode_accept} shows the acceptance gain on math and code, where raising \(K\) from 7 to 15 lengthens \thiswork's acceptance by 1.31/1.19/1.17\(\times\) on \qwen/\glmflash/\glm and \dspark's by 1.27/1.19/1.13\(\times\).
As shown in \cref{fig:k7_analysis}(c), \(K\)=15 is preferable at low load, since its higher acceptance translates into higher user interactivity.
At higher load, however, every speculative baseline pays the drafter overhead of \(K\)=15 and delivers lower total throughput than its \(K\)=7 counterpart at the same user interactivity.

On the other hand, \thiswork jointly schedules draft and verification lengths, so it matches the user interactivity of \(K\)=15 at low load and still achieves the highest throughput at high load.
This is because \thiswork drafts a token only when it is expected to pay off, so a larger \(K\) widens its choices while joint scheduling limits the extra drafter work.
Thus, the main experiments use \(K\)=15, and \thiswork lets serving systems benefit from long drafts with reduced drafter overhead under load.

\section{WCS with Other Drafters Details}
\label{app:orthogonal}

All panels of \cref{fig:wcs_drafters} run \glmflash on ShareGPT-Decode at loads up to 192.
WCS on \dspark uses a per-position pre-draft estimator, which assumes a shorter draft is a prefix of a longer one.
We therefore retrain \dspark with causal attention within the block and train the estimator alongside it, and both \dspark and \dspark+WCS serve this checkpoint.
\mtp and \eagle are autoregressive, so each draft position costs one drafter pass over the whole batch.
Shortening only some requests' drafts saves little time, so WCS instead picks one draft depth per step, up to 5, together with the budget \(B\), using drafter latencies profiled at each depth.
Their pre-draft estimator predicts acceptance at each of the five positions from the target's hidden state at the last accepted token and the anchor embedding.
We train it on a serving log and calibrate it on Alpaca.
After drafting, the budget distributor spends \(B\) using vLLM's online acceptance estimates for these drafters, a per-position logistic model of the drafter's top-token probability that vLLM refits while serving.
In all panels, the gains appear at high load, and the curves converge at low load.

\section{Additional Related Work}
\label{app:related}

\textbf{Parallel speculative decoding.}
Speculative decoding~\cite{leviathan23,chen2023sps} has been strengthened with extra decoding heads~\cite{medusa}, $n$-gram drafting~\cite{lookahead}, feature-conditioned autoregressive drafters~\cite{eagle,eagle2,eagle3}, tree verification~\cite{specinfer,sequoia}, and draft-model distillation~\cite{distillspec,osd}.
Following blockwise parallel decoding~\cite{blockwise}, parallel drafters propose a block in parallel with diffusion language models~\cite{specdiff,specdiff2,diffuspec} or in one pass with lightweight drafters~\cite{dflash,pard}, and later designs restore intra-block dependency~\cite{dspark,dflash2,jetspec}, build draft trees~\cite{ddtree,caddtree}, or distill for higher acceptance~\cite{adaflash}, and these drafter designs can be orthogonal to \thiswork, which decides how much a given parallel drafter drafts and verifies.